\documentstyle[twocolumn]{mn}
\def\beq{\begin{equation}}
\def\eeq{\end{equation}}
\def\bey{\begin{eqnarray}}
\def\eey{\end{eqnarray}}
\def\kms{\mbox{\rm \,km\,s}^{-1}}

\title[]
	{Microlensing of tidal debris on the Magellanic great circle}
\author[]
	{HongSheng Zhao
	\\Max-Planck-Institute f\"ur Astrophysik,
Karl-Schwarzschild-Strasse 1, 85740 Garching, Germany
	\\Sterrewacht Leiden, 
Niels Bohrweg 2, 2333 CA, Leiden, The Netherlands (hsz@strw.LeidenUniv.nl)
\thanks{Present address}}
\date{}
\pagerange{\pageref{firstpage}--\pageref{lastpage}}
\pubyear{1997}

\begin{document}
\maketitle
\label{firstpage}

\begin{abstract}

Increasing evidences suggest that the Galactic halo is lumpy on kpc
scales due to the accretion of at least a dozen small galaxies
(LMC/SMC, Sgr, Fornax etc.).  Faint stars in such lumpy structures can
significant microlense a background star with an optical depth
$10^{-7}-10^{-6}$, which is comparable to the observed value to the
LMC (Alcock et al. 1996c).  The observed several microlensing events
towards the LMC can be explained by a tidal debris tail lying in the
Magellanic plane if the progenitor of the Magellanic Clouds and Stream
and other satellite galaxies in the Magellanic plane has a mass about
twice that of the disc of the LMC.  The LMC stars can either lense
stars in the debris tail behind the LMC, or be lensed by stars in the
part of debris tail in front of the LMC.  The models are consistent
with an elementary particle dominated Galactic halo without massive
compact halo objects (MACHOs).  They also differ from Sahu's (1994)
LMC-self-lensing model by predicting a higher optical depth and event
rate and less concentration of events to the LMC center.

\end{abstract}


\section{Introduction}

Nearly twenty years ago Searle and Zinn (1978) proposed, on the basis
of the spread of horizontal branch morphology of globular clusters and
the lack of a metalicity gradient in the Galactic halo, that the
Galactic halo forms by gradual merging of many infalling sub-Galactic
lumps.  The well-known Magellanic Stream (Mathewson, et al. 1974) and
the newly discovered galaxy in the Sagittarius constellation (Ibata,
Gilmore and Irwin 1994) are the two most impressive signatures printed
by on-going infalls.  The former is a long ribbon of neutral gas
starting from the direction of the Magellanic Clouds, across the south
Galactic pole, occupying $10^o \times 100^o$ of the sky.  The latter is
a faint comoving group of stars in a $5^o \times 20^o$ strip towards
the Galactic center.  If the halo of the Galaxy extends well beyond
the LMC and the SMC, then it would also contain eight other low
luminosity and low surface brightness satellite galaxies (dwarf
galaxies) (Irwin and Hatzidimitriou 1995 and references therein).

Current understanding of the formation of Milky Way size halos based
on collisionless cosmological simulations is that they form with
continuous accretion of numerous smaller halos (White 1996).  On the
order of $\sim 10^{10}M_\odot$ material can be accreted by dynamical
friction with a steady state dark halo (Tremaine 1980, T\'oth and
Ostriker 1992) even after an initial violent merger phase (Toomre and
Toomre 1972).  Material from a minor infall will be confined close to
a specific orbit of a fixed potential.  For an accreted lump on an
orbit with a small peri-galactic radius (within 20 kpc), the tidal
force of the Galaxy becomes strong enough to liberate material from
the lump (Quinn and Goodman 1986, Quinn, Hernquist, Fullagar 1993).
The marginally bound fraction of a small galaxy is peeled off with
each peri-centric passage, and is sent on a gradual leading/trailing
drift nearly along the orbit according to the initial peculiar velocity (Oh,
Lin \& Aarseth, 1995, Johnston, Hernquist \& Bolte 1996).  The tails
generally trace great circles around the host galaxy for nearly
spherical potential in the halo (Lynden-Bell \& Lynden-Bell 1995), and
the tails grow with a rate proportional to the lump's initial velocity
dispersion (Johnston et al. 1996).

Observations on the intrinsic number density of dwarf galaxies in our
halo can set a limit on hierarchical formation models.  The recent
discovery of Sgr has spurred systematic searches for other predicted
small galaxies accreted by the Galaxy in the past one Hubble time
(Johnston et al. 1996, Lynden-Bell and Lynden-Bell 1995, Mateo 1996).
Large tidal tails are generally low surface brightness, kinematically
cold, great circle like moving groups, whose coherent structures are
traceable with halo globular clusters, luminous horizontal branch
stars, giant branch stars or planetary nebulae.  An ideal survey
should combine photometrys, radial velocities and proper motions of
bright objects in a large fraction of the sky.  In the past dwarf
galaxies are mainly found in line of sight directions which have been
frequently studied for other interesting astronomical objects, e.g.,
peculiar stars, halo globular clusters.  This biases against those in
``empty'' or ``boring'' line of sight directions.  The low surface
brightness of dwarf galaxies also makes it more challenging to detect
in line of sight where there are high surface brightness extended
objects, e.g., the Galactic bulge and the Magellanic Clouds.  The
eight known dwarf galaxies and the late surrendipiously discovered Sgr
dwarf galaxy show perhaps more of the need for optimized survey in the
past (ideally a photometric and kinematic survey of luminous tracers
of a large fraction of the sky) than the intrinsic rareness of these
faint structures.  The Sgr, which is about $1/3$ the distance to the
SMC with a comparable total mass at the high end of current estimate
$10^{8-9}M_\odot$ (Ibata, Gilmore, Wyse, \& Suntzeff 1997), has evaded
previous frequent studies of the same low-extinction Galactic bulge
region of the sky, and was found as soon as a kinematic survey of
K-giants at high bulge fields was completed (Ibata et al. 1995).


Current microlensing surveys to the Galactic bulge and towards the
Magellanic Clouds offer good possibility to detect such debris; stars
in the debris are detectable as lenses, amplified sources or bright
variables.  In fact a relative faint part of the Sgr galaxy, which is
roughly behind the Galactic center, was first seen in RR Lyraes in the
microlensing surveys to the bulge (Alard 1996).  A star in Sgr has
also a very high chance being lensed by the stars in the Galactic
bulge and disc, which might be detectable despite the low density of
the sources.  As for surveys towards the Large Magellanic Cloud, 5-8
microlensing events of possiblely LMC sources are observed (Alcock
1996c).  However, possibility of lensing by tidal debris has not been
studied.  The conventional interpretation involves a mathematically
smooth $r^{-2}$ power-law dark halo made of a mixture of massive
compact halo objects (MACHOs) and weakly interacting massive particles
(WIMPs); the MACHO collaboration favors a Galactic isothermal dark
halo with about half of the dynamical mass $2^{+1.2}_{-0.7} \times
10^{11}M_\odot$ out to the distance of the LMC (50 kpc) in
$0.5^{+0.3}_{-0.2} M_\odot$ white dwarf (WD) mass objects and the rest
in distributed form (Alcock et al. 1996a,c and references therein).
Clearly these conclusions hinge on cosmological bases for a smooth
$r^{-2}$ dark halo and the dichotomy of the dark mass (why either
WIMPs or WDs?) in the halo.  Explaining the shortage of lensing events
simply by a (universal?)  MACHO-to-WIMP ratio has the potential
problem of trivializing the (possibly complex) structure in the halo
which we know very little.  Also crucial is a proper estimation of
background non-MACHO events, for example, events due to stellar lenses
in the LMC's bar (Sahu 1994, Bennett et al 1996), and stellar lenses
in clumpy structures in the halo, ranging from globular cluster to
dwarf galaxy sized clumps (Maoz 1994, Metcalf and Silk 1996, Zhao
1996).

In this paper, I show that gravitational microlensing surveys have the
potential of detecting tidal tails in the halo with sizes ranging from
the Sgr to the Magellanic Stream.  In particular I make the possible connection
between hierarchical formation scenario and microlensing events observed 
towards the LMC from ongoing experiment of MACHO
collaboration.  If a small fraction (1/10) of the
Galactic halo's mass inside 16 kpc (the Sgr's distance from the
Galactic center) were accreted from late infall of $10^2$ Sgr size
objects each with a sky angular covering factor $0.25\%$, then there
is a significant probability $\sim 25\%$ of having one towards the
LMC, SMC or any other line of sight.  In reality the amount of late
infall might be limited, at least at small radii,
because high proper motion local halo sample shows
only a weak tail of stars bluer than the bulk of the stellar halo with
$B-V \sim (0.4-1)$ (Unavane, Wyse, \& Gilmore 1996).  Nevertheless 
tidal material surrounding the Magellanic Clouds, and/or
any tidal debris lying on the great circle drawn by the Magellanic Stream,
Ursa Minor and Draco, 
would be consistent with the merging of the
Magellanic Clouds with the Galaxy (Kunkel 1979, Lynden-Bell 1976,
1982).  \S2 estimate the lensing probability (optical depth) towards
the Magellanic Clouds in two configurations of tidal debris: a uniform
grand tidal tail and a short Sgr-like tidal tail on the Magellanic
great circle.  \S3 considers the case that tidal debris is surrounding 
the LMC.  \S4 calculates the event time scales and lens mass function.
\S5 compares several interpretations of the observed events towards
the LMC.  \S6 discusses search techniques.

\section{Lensing by tidal debris on the Magellanic great circle}

\subsection{Evidences for tidal debris}

A yet-to-be-detected strip of faint stars around the Magellanic great
circle has long been predicted in merger models where an ancient
gas-rich lump, supposedly the progenitor of LMC and SMC, was captured
and torn apart by tidal force of the Galaxy (Lynden-Bell 1976, Kunkel
and Demers 1977).  As the lump spirals down the halo of the Galaxy due
to dynamical friction on a massive lump (Tremaine 1976), the
increasing tidal force of the Galaxy, together with a possibly recent
close encounter of the SMC with the LMC, and the ram stripping in the halo,
liberates stars as well as gas from the lump during its last 1-6 orbits.
This merging event might have created several structures which we now
see in the vicinity of the Magellanic great circle: the $100^o$ long
Magellanic Stream (Lin and Lynden-Bell 1982, Murai and Fujimoto 1980,
Lin, Jones, \& Klemola 1995 and references therein), the irregular
Large Cloud, the strongly prolate Small Cloud (Caldwell \& Coulsin 1986), 
the inter-Cloud Bridge with distinct gas lumps (McGee \& Newton 1986) 
and stellar associations (Grondin, Demers \& Kunkel 1992) and
eight known dwarf spheroids scattered on
the Magellanic great circle (Lynden-Bell 1976) or nearby planes
(Kunkel 1979, Lynden-Bell 1982, Lynden-Bell and Lynden-Bell 1995).
Early radial velocity data of the tracers, including a number of
globular clusters and high velocity clouds compiled in Kunkel (1979),
also support this explanation.  A polar orbit would also account for
the nearly polar elongation of the Ursa Minor and Draco galaxies and
their surrounding high velocity clouds; that they are at the almost
opposite direction of the Magellanic Clouds might be an example of a
stable collinear three-body systems in an extended halo (Hunter and
Tremaine 1977).  It is possible that these two dwarf galaxies
are separated from the Clouds much early on.  The material now in the
inter-Cloud Bridge and the Magellanic Stream might be
mostly due to a recent encounter of the Clouds.

The Magellanic Clouds should be shrouded with a faint inhomogeneous
band of liberated material, gas or stars as a result of the merger
event which dynamically links together several known structures on the
Magellanic great circle.  
Numerical simulations shows such a situation
would be very natural; for example, the tidal material between 30-70
kpc in Fig.11 of Gardiner et al. (1994), sandwiching the Large Cloud.
Observationally the surrounding debris is perhaps manifested by the
gas (McGee \& Newton 1986) and star clusters (Irwin 1991 and
references therein) bridging the Large Cloud and the Small Cloud.
Some five stellar associations in the region also show a distance
spread as much as $17$ kpc (Grondin, Demers \& Kunkel 1992).  The huge
depth of the SMC ($\sim 10$ kpc) is also suggestive of a recent
encounter with the LMC; see the comparision of SMC Cepheid distances
(Caldwell \& Coulson 1986) with simulation in Fig.13 of Gardiner et
al. (1994).  

{\it As long as the above picture is qualitatively correct, it is
inevitable that stars in the surrounding debris will have a fair
chance to microlense or be microlensed by stars in the LMC and SMC,
contributing a few events in the current microlensing surveys towards
the Magellanic Clouds.}

Have past observations of the LMC pretty much ruled out any large
moving group right in front of the LMC?  Probably not, if the two
differ in distance modulus by only $0.5$ to $0.9$ magnitude
(corresponding to the foreground material at about 4/5 to 2/3 of the
distance to the LMC), and if the contrast in the column density
between the low surface brightness lump and the LMC is around 100.
The Sgr dwarf, for example, has a density about $1$ per square arcmin
for horizontal branch stars (Ibata et al. 1995), which is roughly 3\%
of the density of bulge stars at the same field (Alard 1996, Alcock et
al. 1996b).  A CCD observation of a field with one sq. armin area of
the LMC might yield an overlapping CM diagram of $10^4$ stars from the
LMC and $10^2$ stars from the foreground material, both of mixed
stellar populations due to strong age and metallicity spread.  The
half a magnitude difference and the weak contrast might be considered
insignificant given internal extinction of the LMC (can be as much as
1.5 mag. at the LMC bar, Sahu 1994 and references therein) and small
number statistics.  The low contrast, the faintness, the large size,
and the very irregular morphology of the LMC also conspire to make
detection of any tidal tails in small degree size photographic plates
of the LMC difficult.  The Sgr horizontal branch stars are spread out
over many degrees of the sky with a density about $1$ per square
arcmin and $\sim 19$ magnitude in $V$.  Stellar populations of the
Galactic bulge have been studied for 50 years since Baade's (1946)
pioneering discovery of variable stars in the direction, the recently
found Sgr shows the limitations of previous observations.

It is highly promising to search for RR Lyraes, 
and other variable populations of the debris in the variable star
catalogues of current microlensing surveys.  A weak indication of
foreground material is already seen in earlier surveys of variable
stars of the region.  Payne-Gaposchkin (1971) published a survey of
variable stars in the direction of the LMC, listing 29 short-period
variables as foreground RR Lyraes with distances between 5 and 25 kpc.
Interestingly 9 of these cluster at a narrow distance modulus range
$16-17$ mag, which corresponds to a line-of-sight distance about 16 to
25 kpc.  The follow-up photometric and spectroscopic studies by
Connolly (1985) and Smith (1985) confirmed that most of these
short-period variables are indeed foreground metal poor RR Lyraes with
a radial velocity different from LMC RR Lyraes.  Since only about 1 RR
Lyrae between 5 to 16 kpc, and also between 16 to 25 kpc towards the
LMC are expected from a smooth $r^{-3.5}$ or $r^{-4}$ density law for
the RR Lyraes in the stellar halo (Saha 1985), the 9 RR Lyraes with
distance modulus $16-17$ mag. seem to trace a local over-density
region.

Given the above lines of evidence for tidal debris and the important
implications on the nature of the dark matter, it becomes highly
interesting to examine the lensing optical depth of the debris in some
detail.

\subsection{Observed optical depth to the LMC}

The observed optical depth to the LMC is still uncertain.  The number
of claimed microlensing events from the MACHO and EROS experiments
has been fluctuating between 1 to
half a dozen.  Based on 5-8 events which pass their recent
selection criteria the MACHO team 
estimated an optical depth towards the LMC
\beq \label{obs}
\tau_{obs} = 0.17 ^{+0.09}_{-0.05}\times 10^{-6};
\eeq
a somewhat higher value is also given in a more recent 
estimation (Alcock et al. 1996c).

\subsection{Optical depth and the mass of a grand Magellanic tidal tail}

Theoretically it is still premature to predict a precise range of
microlensing optical depth accountable by the debris of the old
Magellanic system without a detailed N-body modeling of the morphology
and the clumpyness of the tail and the total mass of the system.
Nevertheless some insights to the problem can be obtained for some
greatly simplified cases.  The morphologies which I will assign
to the debris tail should be treated as toy models only.  They clearly 
cannot match the variety and complexity of tidal tails frequently
seen in mergers and N-body simulations (e.g. Toomre \& Toomre 1972).

This section deals with a grand {\sl
uniform} tidal tail due to a galaxy slightly more massive than the
Magellanic Clouds.  The clumpyness of the grand tidal tail is modelled
in \S3 by {\sl adding} a small kpc size tidal tail of a disrupted
dwarf galaxy on top of the uniform grand tidal tail.  In both cases
only debris in the front side of the LMC is considered.  The consequences of
tidal debris surrounding the LMC or {\sl behind} the LMC are studied in \S4.

It is well-known that the optical depth towards a source at distance $D_s$ is
given by (Paczy\'nski 1986)
\beq\label{basic}
\tau ={4 \pi G \over c^2} D \Sigma = {4 \pi G \over c^2} \int_0^{D_s}\!\! (1-x) D_l \rho(D_l) dD_l,~~x\equiv{D_l \over D_s},
\eeq
where 
\beq
D\equiv D_s x(1-x)
\eeq
and $\rho(D_l)$ are the effective distance and 
the density of the lens at distance $D_l$,
$\Sigma$ is the surface density of lenses.
So the lensing optical depth by
the debris tail depends on geometry of our line
of sight to the LMC with the debris plane, and the density
distribution along the tail.

As an order of magnitude estimation, it is reasonable to assume the
debris tail has a uniform torus with a cross section $\pi b^2$, a
length $L$, a total mass $M$, and a uniform density $\rho={M \over \pi
b^2 L}$.  This picture is motivated by N-body simulations of accretion
of a dwarf galaxy, where the dwarf galaxy is shown to be disrupted
into a ring like distribution (Quinn and Goodman 1986, Quinn,
Hernquist, Fullagar 1993).  Assume that the Sun is sufficiently close
to the Galactic center, so that our line of sight to the LMC is
parallel to the debris plane, and in fact, inside the debris central
plane within $\pm b$.  For the time being 
assume the debris tail is sufficiently wrapped
around the sky, so that the optical depth is nearly the same around
all line-of-sight directions within $\pm b$ of the central plane of the debris.
Take the average of these line of sight directions, I have 
\bey
\left<\tau \right> &\approx & {4 \pi G \over c^2} \int_{-b}^{b} {dz \over 2 b}
\int_{-\pi}^{\pi} {d\theta \over 2 \pi} \int_0^{D_s} dD_l D_l
\rho(D_l)\\
&=&{G M \over c^2 b}, ~~~M=\int dr^3 \rho
\eey
where I have made the approximation that $(1-x) \approx 1$ in
eq.~\ref{basic}.

More generally, I find
\beq\label{debris}
\tau = {G M \over c^2 b} \xi = 0.17 \times 10^{-6} 
{M \over 1.5 \times 10^{10} M_\odot} {8 {\rm kpc} \over 2b} \xi,
\eeq
where $\xi$ is a dimensionless quantity of order unity, which 
takes into account of the $(1-x)$ factor, and 
the fact that the optical depth can be a strong function of 
directions for
a short debris tail which wraps around the sky only once or less.

For the validity of the plane parallel approximation, the thickness of
the debris plane, $2 b$, must be at least comparable to the Sun's
distance from the Galactic center $R_0$, so $2 b \ge R_0= 8$ kpc,
which translates to about $9^o$ at the LMC's distance.  This is
consistent with the width of the Magellanic Stream $\sim 10^o$.  The
luminous disc of the LMC has a mass of about $10^{10}M_{\odot}$.  So
eq.~\ref{debris} and~\ref{obs} together imply that {\sl a low surface
brightness debris tail with mass in stellar objects comparable to the
disc of the LMC is needed to explain the observed microlensing optical
depth}.  If the ancient Magellanic system (Lynden-Bell 1976, Kunkel
and Demers 1977) has the combined mass of the debris inferred here,
the Magellanic Clouds, and several satellite galaxies and high
velocity clouds on nearby great circles, I estimate its dynamical mass
between $(2-4)\times 10^{10}M_\odot$.  

A lump with mass about twice
that of the Magellanic Clouds would reduce its orbital radius (with
same eccentricity) by half in half a Hubble time due to dynamical
friction with an extended halo of the Galaxy (Tremaine 1976, Murai and
Fujimoto 1980); the de-acceleration is proportional to the satellite's
mass.  A possible scenario for the structures around the Magellanic
great circle is that the ancient Magellanic system on a polar
eccentric orbit was probably disrupted during the first pericenter
passages around $50$ kpc to liberate some low mass dwarf galaxies such as
Ursa Minor and Draco from
the binary Magellanic Clouds.  The later continues its doomed course
deeper in the halo because of still strong dynamical friction.  The
SMC was disrupted by a perhaps recent close approach to the LMC, and
the material is then liberated by the halo tidal force to form the inter-Cloud
Bridge and the giant Magellanic Stream (Gardiner et al. 1994, 1996).  
Ram pressure stripping due to a possible extended gaseous halo may have also
played a role for the Stream (Moore \& Davis 1994).

The above optical depth 
estimation applies also to {\it a localized debris distribution
surrounding the LMC}, such as the inter-Cloud Bridge, perhaps as a
result of the tidal shock in LMC's recent close encounter with the
SMC.  For simplicity, one might model the debris
as a faint prolate-shaped distribution pointing in the
direction of the LMC's space velocity.  In this case, the $\xi$ factor
in eq.~\ref{debris} is reduced by a factor $(1-x) \sim {b \over D_s}$
because the lens and the source are very close, but is enhanced by a
factor ${2\pi D_s\over b}$ because lensing is concentrated to a small
angular region of the sky.  So the two effects balance out, and $\xi$
is still of order unity.  Again debris of the mass of the order
$10^{10}M_\odot$ is necessary; the exact result depends on the axis
ratio and the angle of the prolate-shaped distribution with our line
of sight, similar to the self-lensing of the Galactic bar (Zhao and Mao 1996).

If the debris tail is both short and far from the LMC, then the
optical depth gains by a factor $\xi \sim {2 \pi D_s \over b}$.  In
this case a tidal tail with mass $10^{8-9}M_\odot$ is enough to
explain the observed optical depth.

\subsection{Optical depth of faint stars in a Sgr size galaxy}

The calculation of the previous section assumes a homogeneous tidal
tail on the Magellanic great circle, while a lumpy distribution is
more plausible.  If the ancient Magellanic lump had a few dwarf
galaxies, those less dense than Ursa Minor and Draco might have been
disintegrated into small tidal debris tails at their peri-galactic
passes.  The most efficient way to produce microlensing towards the
LMC is to have one of these small tidal tails right in front of the
LMC.  Although fairly contrived, this particular model has a few easy
to test predictions.

Let us estimate the microlensing optical depth due to lenses
in a Sgr-like dwarf galaxy.  The formalism is the same as used in deriving
the optical depth of the dark halo (Paczy\'nski 1986)
except that the situation here is simpler since all the lenses are at nearly
the same distance.  Lacking the knowledge of the real
distribution of these dwarf galaxies, I will simply ``move'' the Sgr dwarf
galaxy in front of the LMC, and compute the microlensing probability
of source stars in the LMC.

For the lenses in the dwarf galaxy between us and the LMC, the optical depth is
(cf.~\ref{basic}) 
\beq\label{dg}
\tau_{dg}= 0.17 \times 10^{-6} {\Sigma \over 20 M_\odot\mbox{\rm pc}^{-2} } {D \over 12 \mbox{\rm kpc} },
\eeq
where I have scaled the physical quantities with characteristic values.

$\Sigma$ is related to the surface brightness $\mu$ by
$\Sigma=\Upsilon \mu$, where $\Upsilon$ is the mass-to-light ratio,
but more precisely, the ratio of total mass in stars and other compact
objects of the dwarf galaxy to the total stellar light of the dwarf
galaxy.  For the Sgr, the surface brightness of the dwarf galaxy
($\mu$) is $4 L_\odot pc^{-2}$ near the nucleus and is decreased to
about $1.5 L_\odot pc^{-2}$ at $10^o$ from the nucleus (Ibata et
al. 1995, Mateo et al. 1995, 1996).  Dwarf galaxies are generally
dominated by dark matter from the core to the tidal radius with the
total mass to total light ratio $M/L$ probably in the range
$5-200M_\odot L^{-1}_\odot$ (Irwin and Hatzidimitriou 1995).  Ibata et
al.  (1997) estimate that $M/L \sim 100M_\odot L^{-1}_\odot$ for the
Sgr.  To make a conservative estimation for the lensing optical depth,
I assume $\Upsilon=10 M_\odot L^{-1}_\odot$, which is reasonable for
an old population.  This way the dwarf is dominated by a non-baryonic
halo, and only a small fraction of the dynamical mass is in faint
stars, which can lense.

The distance to the Sgr dwarf galaxy ($D_d$) is about ${1 \over 2}$
of the distance to the LMC ($D_s$), so $D_s \approx 2 D_d \approx 50$
kpc, and $D \approx 12$ kpc.  Assume 
$\Upsilon \approx 10 M_\odot L^{-1}_\odot$, 
and $\mu=(1.5-4)L_\odot {\rm pc}^{-2}$.
I find the optical depth
\beq\label{comp}
\tau_{dg} = \left({\mu \over 2 L_\odot {\rm pc}^{-2} }\right) \tau_{obs} = (0.8-2) \times \tau_{obs},
\eeq
is {\sl comparable to the observed value to the LMC} $\tau_{obs}$ as given in
eq.~\ref{obs}.

This conclusion is insensitive to the exact value of the dwarf's
distance, because 
\beq
\tau_{dg} \propto D \propto x (1 - x),~~~x={D_d \over D_s},
\eeq
so $\tau_{dg}$ has a very broad peak at $x=1/2$.  
Compared to the optimal half-way position which is adopted in
eq.~\ref{comp}, the optical depth reduces only by $11\%$ if the dwarf
galaxy is at $2/3$ or $1/3$ of our distance to the LMC, by $25\%$ at
$3/4$ or $1/4$, and by $36\%$ at $4/5$ or $1/5$.  In the case
that the tidal material is within 5 kpc to the LMC, the debris
will be mixed with the material surrounding the LMC.

The optical depth is also insensitive to a small misalignment of the
dwarf nucleus with the LMC.  The Sgr has a very shallow major axis
gradient of the surface brightness: I estimate an e-folding angular
size from the nucleus about $10^o$.  A tidal tail with such size is
big enough to ``cover'' the sky area of the LMC.  

Nevertheless it is unlikely for a Sgr-like tidal tail to cover both
the LMC and the SMC, which are spaced more than twenty degrees apart.
A comparable lensing optical depth to both the LMC and the SMC could
rule out such a small tidal tail.  By the same line of argument, a
high lensing rate to the Andromeda galaxy, which is off the Magellanic
great circle, would lend support to the explanation of the LMC events
being due to a smooth distribution of MACHOs rather than a massive
grand tidal tail on the Magellanic great circle.  On the other hand,
any sharp gradient of the event rate across the solid angle of the
Andromeda galaxy would be indications for clumpyness in its own halo.

\section{Stars on tidal debris as sources for lensing}

Stars in the debris can also be sources for lensing.
If the Magellanic Clouds are shrouded with tidal debris, then stars of
the Clouds can also lense those in the background tidal debris.  It is
the opposite situation compared to the situation where foreground
debris stars lense background Magellanic Cloud stars, and has the
advantage of hiding the debris easily.  Both are very different from
Sahu's suggestion of LMC self-lensing, as the debris can be quite a
distance (several kpc) in front or behind the LMC, so the lensing
optical depth can be much higher; the depth of the disturbed SMC and
and the inter-Cloud Bridge argues for debris at distances around
40-70 kpc from us (cf. Figure 11 of Gardiner et al. 1994).  When the
debris is at the back of the LMC the source density will be much lower
(by a factor $\mu_{db}/\mu_{LMC} \sim 0.1$ based on the density of the
RR Lyraes in Sgr and LMC), but the lens density (hence the optical
depth) increases by the same factor.  As a result the microlensing
event {\sl rate} will be comparable for both cases, because the event
rate is proportional to the product of the number density of stars in
the debris and those in the LMC in both cases.

More rigorously, for a survey with $N_s$
background source stars and $N_f$ foreground point masses in the
survey solid angle $\Omega$, the expected number of events is given by
\beq
N_{e} = \epsilon N_s N_f \Omega_m /\Omega
\eeq
where $\epsilon$ is the survey efficiency,
and $\Omega_m$ is the solid angle significantly microlensed 
by a foreground moving point mass $m$ during the survey,
which is the angular area of the Einstein ring plus a rectangular area 
which is swept by the moving ring in the survey time $T$,
so
\beq
\Omega_m=\pi \theta_E^2 + (2 \theta_E) (\left|\vec{\omega}_{ls}\right| T),
\eeq
where 
\beq
2 \theta_E= 4 \left[ {G m \over c^2 }
\left( {1 \over D_l} - {1 \over D_s} \right) \right]^{1 \over 2},
\eeq
is the angular diameter of the Einstein ring for 
significant microlensing amplification (0.34 magnitude) and 
\beq
\vec{\omega}_{ls} = {\vec{v}_{lo} \over D_l} - {\vec{v}_{so} \over D_s} 
\eeq
is the relative proper motion rate of the source and 
the lens, and $\vec{v}_{so}$ and $\vec{v}_{lo}$ are the transverse 
velocities of 
the source and lens with respect to the observer.
$N_s$ and $N_l$ are simply related to 
the surface mass densities $\Sigma_s$ and $\Sigma$ for the background sources
(with mass $m_s$) and the foreground point masses multiplied by the areas,
\beq
N_s = \left({\Sigma_s \over m_s} \right) \left(\Omega D_s^2\right),~~~N_l = \left({\Sigma \over m} \right) \left(\Omega D_l^2\right).
\eeq
As a result, in the limit that the survey time is much longer than any single event,
\beq\label{ne}
N_e \approx K T \Omega \Sigma_s \Sigma,
\eeq
where
\beq
K=4 G^{1 \over 2} c^{-1} m^{-{1 \over 2}} m_s^{-1} D_s^{2.5} \sqrt{x (1-x)}
\left| \vec{V} \right|,
\eeq
and
\beq
\vec{V}=\vec{v}_{lo} - \vec{v}_{so} x.
\eeq

Clearly $N_e$ depends mainly on the product of the surface density
in the background and in the foreground, so moving the debris from in
front of the LMC to behind the LMC gives comparable number of events.
For fixed $x$ (say $x=1/2-5/6$), the $D_s^{2.5}$ dependence in fact
favors a far away source by a factor
$\left(D_s/D_{LMC}\right)^{2.5}=x^{-2.5}=(1.6-5.7)$.  However, the
efficiency $\epsilon$ biases against faraway sources with perhaps
$\epsilon \propto D_s^{2 \beta}$ due to a detection limit (Kiraga and
Paczy\'nski 1994), so that $K \propto D_s^{2.5-2\beta}$ is insensitive
to source distance if $\beta$ is between $1$ and $1.5$.  The
efficiency also depends on the amount of dust in front of the source.
The number of observable sources behind the LMC is reduced to a factor
$1/y \approx 1/3$ (Sahu 1994) of intrinsic value due to screen-like
dust extinction.  The number of observable LMC sources is reduced to a
factor $(1-1/y)/\log y \approx 0.6$ due to self-extinction.  The
values of $K$  in both cases are approximately the same within a factor of two,
insensitive to the values of $y$ and $\beta$.  So the number of events
expected for a background debris tail is roughly the same as for the
foreground debris tail.
\footnote{An effect left out here is that $\epsilon$ is also a
function of amount of source blending, which is important for crowded
fields in the LMC}.

\section{Lens mass function and event time-scales}

\subsection{Lens mass function}

The Einstein diameter crossing time (Paczy\'nski 1986)
as a function of the mass of the lens $m$, the lens's relative 
velocity transverse to the observer-source line of sight 
$\left| \vec{V} \right|$
and the effective distance $D$ is given by
\bey
2 t_E &\equiv& {2 \theta_E \over \left| \vec{\omega}_{ls} \right|} \\
&\approx& 34 \mbox{\rm d.} \left({m \over 0.1 M_\odot}\right)^{1 \over 2} 
\left({D \over 12 kpc}\right)^{1 \over 2} 
\left({ \left| \vec{V} \right| \over 300 \kms}\right)^{-1}.
\eey
where I have scaled the quantities with characteristic values.

The 6 events recently reported by the MACHO collaboration have
$2\times t_E=34-114$ days if excluding one binary event and one low
confidence event (Alcock et al. 1996c).  If the lenses of the LMC
events were in a foreground dwarf galaxy rather than in the isothermal MACHO
halo, then the spread of the lens's distance and velocity will
be negligibly small, so the range in the duration of observed
events is purely due to the mass function of the lenses.

If the dwarf galaxy is behind the LMC, and acts as the source instead
of the lens, then the distribution of the effective distance and
relative velocity will still be narrow.  In both cases the roughly
factor of $3$ range in the observed $t_E$ translates to about a factor
of $10$ in the lens mass $m$.  The actual median mass depends
sensitively on the proper motion of the dwarf galaxy and somewhat on
its distance; a mass range between $0.1M_\odot$ to $1M_\odot$ is
certainly {\it a} plausible solution for some combinations of
distances and velocities.

\subsection{Tidal debris}

Most likely either the lenses or the sources are part of the tidal
debris in the Magellanic great circle.  They could be in the form a
grand smooth tidal tail or several smaller tidal tails from disrupted
dwarf galaxies in the plane.  In all these cases several observable
properties can be predicted.

The tidal debris has a transverse velocity
proportional to that of the center of mass of the Magellanic Clouds
\beq
\vec{v}_{td}=x^{-1}_{td} \vec{v}_{LMC},
\eeq
because they share the same great circle orbit, hence have the same specific
angular momentum; I have neglected the small offset of the Sun from
the Galactic center and put the LMC at the center of mass of the 
Magellanic Clouds.  The LMC has a proper motion of $1.2 {\rm mas yr}^{-1}$
(Jones, Klemola, Lin 1994), 
which translates to a Galactocentric transverse velocity 
$\left|\vec{v}_{LMC}\right| \approx 200 {\rm km s}^{-1}$ in direction
leading the Magellanic Stream.
So the relative proper motion between 
the tidal debris and the LMC is
\beq
\vec{\omega} = \left(1-x^{-2}_{td}\right) \vec{\omega}_{LMC}, ~~~\vec{\omega}_{LMC} = {\vec{v}_{LMC} \over D_{LMC} },
\eeq
where I have left out the parallax due to the Sun's motion.
So when peculiar 
motions of stars in the Magellanic Clouds and tidal debris are small
compared to $\vec{\omega}$, then  
the model predicts that the lens-source proper motion vector 
$\vec{\omega}_{ls}$ will be
approximately $\vec{\omega}$, which is
parallel to the great circle of the Magellanic Stream, with an amplitude
\beq
\left| \vec{\omega}_{ls} \right| 
\approx \left| \vec{\omega} \right| 
\approx \left| \vec{\omega}_{LMC} \right| \left|1-x^{-2}_{td}\right|.
\eeq
The prediction can be compared with parallax measurements for individual
microlensing events, which can determine the amplitude and  direction of
the reduced velocity $\vec{V}/(1-x)$ (Gould 1994).  The amplitude
is predicted to be 
$\left| \vec{\omega}_{ls} \right| D_s x/(1-x)
\approx 200 {\rm km s}^{-1} (1+x^{-1}_{td})$.
However, the prediction is subject to large uncertainty when
the tidal debris is very close to the LMC, $x_{td}$ between $0.9-1.1$,
so that 
the systematic velocity between the debris and the LMC is comparable
to that of the rotation velocity of the LMC stars ($\sim 70$ km/s).
When the debris tail is thick, the predicted amplitude (not direction)
also varies for each event with
the detailed location ($x_{td}$) of the lens or source.

The event time scale can also be predicted given a detailed kinematic
model.  In the current simple kinematic model $t_E$ is a function of $x_{td}$.
\beq
2 t_E 
\approx 34 \mbox{\rm d.} \left({m \over 0.1 M_\odot}\right)^{1 \over 2} g(x_{td})
\eeq
where
\beq
g(x_{td}) =
\left( { \left| \vec{\omega}_{ls} \right| \over 1.2 {\rm mas yr}^{-1} }\right)^{-1} \sqrt{1-Min[x_{td},x^{-1}_{td}] \over 0.25}.
\eeq
For typical foreground debris distances $x_{td}$ from $2/3$ to $4/5$,
$g(x_{td})$ is between $2$ to $2.5$; for typical background debris
distances $x^{-1}_{td}$ from $2/3$ to $4/5$,
$g(x_{td})$ is between $0.9$ to $1.6$.  
So in both cases, a typical lens mass (in the foreground debris or in the LMC)
of $0.1M_\odot$ is roughly consistent with the observed events.

With the increasing number of events, and a self-consistent N-body
simulation of the infall of the Magellanic Clouds, which could predict
the lens and source kinematics, it would be possible to test whether
the derived lens mass function is consistent the luminosity function
of the Large or Small Cloud.  {\it If the lenses were from stars liberated
from the Large or Small Clouds, they should have the same mass
function as their parent galaxies.}

Bennett et al.  (1996) found a caustic crossing binary event in the
MACHO LMC data.  Assuming the caustics are resolved by the finite size
of a late A-type source star near the LMC disc, they derive an
extremely low lens-source relative proper motion speed ($\sim 20$
km/s).  Such small relative velocity is very unlikely for a halo lens,
but is possible if either the source or the binary lens is from the
tidal debris within 2.5 kpc to the LMC, $|1-x_{td}|<0.05$.  In this
case the relative systematic velocity between the debris and the LMC
is low, $\le 30$ km/s, so the velocity dispersion ($\sim 20$ km/s) of
the LMC stars becomes important, which can reduce the lens-source speed
to the observed value.  

This explanation is also consistent with most of the observed events
being from the tidal debris.  Caustic crossing binary events are
intrinsically rare, in fact less than a handful of caustic crossing
binary events are found among more than one hundred events towards the
Galactic bulge and the LMC.  Bennett et al. found that it is difficult
to explain all LMC events by the LMC self-lensing due to its low
optical depth if the LMC disc is close to face-on (Gould 1995).  But
the observed 5-10 ordinary events and one caustic event is consistent
with them being from sources or lenses in the tidal debris with a few
kpc of the LMC.  It would be interesting to see whether we will detect
another such LMC comoving binary event when the number of LMC events
doubles in the coming year.

\section{Possible interpretations of the LMC lensing events}

There are a few simple interpretations of the non-zero but low optical
depth towards the LMC.

(1) The dark halo is in distributed particles, and either (a) events
due to self-lensing of LMC stars (Sahu 1994) are underestimated, or
(b) background events due to a non-uniform stellar halo of the Galaxy
are underestimated.

(2) There are numerous MACHOs in the halo, but either (a) the entire
halo is made of MACHOs but with a density distribution varying
markedly with line-of-sight directions, or (b) the halo indeed has a
smooth $r^{-2}$ distribution but with a constant mix of white dwarf
like objects and weakly interacting massive particles (Alcock et al. 1996c).



Even a small mass fluctuation in the halo can greatly influence
conclusions on the amount of massive compact objects in the halo.  The
projected mass density of {\it stars} in a dwarf galaxy, $\sim \left(2
\Upsilon\right) \sim 20 M_\odot pc^{-2}$, is comparable to an
isothermal halo, $\sim \left(V_{cir}^2 f_{MACHO}\right) \left( 4 \pi G
r\right)^{-1} \sim \left(100f_{MACHO}\right) \sim 30 M_\odot pc^{-2}$,
where $f_{MACHO}$ is the fraction of the halo mass in MACHOs, and $r
\sim 8$kpc is the typical distance to the MACHOs, and $V_{cir} \sim
220 \kms$ is the amplitude of the gas rotation curve of the Galaxy.
So faint stars in a $10^{8-9} M_\odot$ satellite galaxy can produce an
optical depth (cf. eq.~\ref{dg} and~\ref{comp}) comparable to that of
a $2\times 10^{11}M_\odot$ isothermal halo of WDs: $\tau_{dh} \sim
f_{MACHO} {V^2_{cir} \over c^2} \sim 0.15 \times 10^{-6}$ (Alcock et
al. 1996a).


If one relaxes the model assumption to allow MACHOs (such as stellar
remnants, brown dwarfs and Jupiter-mass objects) to dominate in both dwarf
galaxies and the Galactic halo (as in case 2a), then the optical depth
of the dwarf can increase by one order of magnitude $ \tau_{dg} \sim
\tau_{obs} {M/L \over 10 M_\odot/L_\odot} \sim 10 \tau_{obs} $
(cf. eq.~\ref{dg} and~\ref{comp}), given the high $M/L \sim 100$ of
dwarfs.  Now suppose the MACHO-dominated dwarf galaxies which come
into merge with the halo has a mass spectrum ${dN \over d\log M} =
N_0 \left(M/10^{10} M_\odot\right)^{-1}$ with mass $M$ from
$3 \times 10^{6-10}M_\odot$, and they are now in various stage of disruption in
the halo; some cover as much as $0.2\%$ of the full sky solid
angle per tidal tail.
Looking through this
clumpy halo the optical depth (and to some extent the event duration)
will be a wildly oscillating function of line-of-sight directions on
scales of several degrees.  In this picture one would need to observe
at least several directions to estimate the average optical depth.
One can not argue any single line of sight, say the LMC, is typical or
not.  It becomes problematic to constrain the spatial distribution and
spectrum of MACHOs with the observed microlensing events in the LMC
direction.


The LMC self-lensing (Sahu 1994) is efficient when the ``depth'' of
the LMC is big.  In this aspect the SMC has a more favorable geometry
than the LMC, the former being a very elongated bar pointing close to
us with a depth about $10$ kpc (Caldwell \& Coulson 1986).  The event
rate is only limited by the source density, which is much lower than
the LMC.  I expect comparable number of events from SMC self-lensing
and lensing by halo dwarf galaxy or MACHOs.  I also estimate
comparable event rate (roughly one event per year for MACHO
experiment) coming from LMC self-lensing, SMC self-lensing, and the
lensing of Sgr by foreground bulge stars.

It is possible to test whether the lenses are on the LMC.  If sources
are the far side of the LMC or behind the LMC, one expects the lensed
sources to be systematically shifted towards fainter and redder part
of the CM diagram than the average LMC stars in the survey due to
distance and extinction.  The event rate is a function of
line-of-sight position, proportional to the surface density of
foreground lenses and that of background sources (cf. eq.~\ref{ne}).
The gradient of the surface density of an isothermal MACHO halo or
tidal tails is generally shallow, the projected density of the LMC is
peaked to its central bar.  Color-Magnitude diagram and spatial
information of the lensed sources can be used to test various
distributions of the sources and lenses.  The event distribution
(spatial, time, color-magnitude) to the SMC and the M31 are also
indicators to differentiate the models.  

Event rate towards the M31 depends on the covering factor of any tidal
debris in its halo and any chance intervening tidal material in our
Galactic halo.  No strong conclusions can yet be drawn from the
current handful of events (Alcock et al. 1996c, Ansari et al. 1996,
Crotts \& Tomaney 1996).

\section{Searching for fossils of past mergers}

The north and south Galactic poles are two good directions to search
for tidal material because polar orbits have zero angular momentum
with respect to the rotation axis of the Galactic disc, and are likely
populated by infalling satellites.  Three postulated streams, the
Magellanic-Draco-Ursa Minor stream, the Fornax-Leo(I \&
II)-Sculptor-Sextans stream, and the Sgr stream are all on polar great
circles.  Stars in a stream are distinguishable from field stars by
their narrow distribution in distance modulus, radial velocity, and
proper motion.  

The studies should combine photometric surevys with kinematics.  Dwarf
galaxies and the Magellanic Clouds are massive enough to contain
globular clusters and planetary nebulae, and they are also rich in
metal poor RR Lyraes and carbon stars.  All these have played
important roles in discovering the Sgr galaxy.  Sky positions and sometimes
distances of these luminous tracers can be used to map the streams in
the halo.  Several methods for interpreting the data have been
developed by exploiting the fact that the tidal tails at large radius
in the halo nearly trace great circles (Lynden-Bell and Lynden-Bell
1995, Johnston et al. 1996).

Since pure photometric surveys are generally inconclusive, kinematic
confirmation is crucial.  The tangential proper motion of a stream
generally has a large positive or negative value ($\sim \pm 200$ km/s)
as the stars in a stream are bunched in phase as well as in orbit;
Majewski, Munn, Hawley (1994) have applied this method to identify a
moving group towards the north Galactic pole.  When an extended region
is studied, a sinusoidal variation of the radial velocity with the
angular position along the orbit is also detectable (Kunkel 1979,
Johnston et al. 1996).  Such distributions are markedly different from
that of a pressure-supported stellar halo.


I thank Simon White and David Spergel for encouragements and Tim de
Zeeuw for a careful reading of the manuscript.  I am obliged to a
discussion with Mario Mateo and Lin Yan on observations. 

{}

\bsp
\label{lastpage}

\begin{thebibliography}{}

\bibitem{} Alard, C. et al. 1996, ApJ, 458, L17
\bibitem{} Alcock, C. et al. 1996a, ApJ, 461, 84
\bibitem{} Alcock, C. et al. 1996b, preprint (astro-ph/9605148)  
\bibitem{} Alcock, C. et al. 1996c, preprint (astro-ph/9606165)  
\bibitem{} Ansari, R. et al. 1996, A\&A, 314, 94 
\bibitem{} Baade, W. 1946, PASP, 58, 249
\bibitem{} Bennett, D. et al. 1996, preprint (astro-ph/9606012) 
\bibitem{} Caldwell J.A.R. \& Coulson I.M. 1986, MNRAS, 218, 223
\bibitem{} Connolly,  L. P. 1985, ApJ, 299, 728
\bibitem{} Crotts, A. \& Tomaney, A. 1996, ApJ 473, L87
\bibitem{} Gardiner, L.T., \& Noguchi, M. 1996, MNRAS, 278, 191
\bibitem{} Gardiner, L.T., Sawa, T., Fujimoto, M. 1994, MNRAS, 266, 567
\bibitem{} Gould, A. 1994, ApJ, 421, L75
\bibitem{} Gould, A. 1995, ApJ, 441, 77
\bibitem{} Grondin L., Demers S. \& Kunkel W.E. 1992, AJ, 103, 1234
\bibitem{} Hernquist, L. \& Mihos, J.C. 1995, ApJ, 448,41 
\bibitem{} Hunter, C. \& Tremaine, S. 1977, AJ, 82, 262
\bibitem{} Ibata, R., Gilmore, G., \& Irwin, M. 1994a, Nature, 370, 194
\bibitem{} Ibata, R., Gilmore, G., \& Irwin, M. 1995, MNRAS, 277, 7811
\bibitem{} Ibata, R., Wyse, R., Gilmore, G., \& Irwin, M., Suntzeff, N. 1997, AJ, 113, 634
\bibitem{} Irwin, M. 1991, in IAU Symp. 148 on ``The Magellanic Clouds'' eds. R. Haynes and D. Milne (Dordrecht: Kluwer), p453
\bibitem{} Irwin, M., Hatzidimitriou, D. 1995, MNRAS, 277, 1354
\bibitem{} Johnston K., Hernquist L., \& Bolte M. 1996, ApJ, 465, 278
\bibitem{} Jones, B.F., Klemola, A.R., \& Lin, D.N.C., 1994, AJ, 107, 1333
\bibitem{} Kiraga, M. \& Paczy\'nski, B. 1994, ApJ,  430, L101
\bibitem{} Kunkel, W.E. 1979, ApJ, 228, 718 
\bibitem{} Kunkel, W.E. and Demers, S. 1977 
\bibitem{} Lin D.N.C. \& Lynden-Bell D. 1982, MNRAS, 198, 707
\bibitem{} Lin D.N.C., Jones B.F.,\& Klemola A.R. 1995, ApJ, 439, 652
\bibitem{} Lynden-Bell, D. 1976, MNRAS, 174, 695 
\bibitem{} Lynden-Bell, D. 1982, Observatory, 102, 202 
\bibitem{} Lynden-Bell, D. \& Lynden-Bell, R.M. 1995, MNRAS, 275, 429 
\bibitem{} Majewski, S.R., Munn, J.A., \& Hawley, S.L. 1994, 427, L37 
\bibitem{} Maoz, E. 1994, ApJ, 428, L5
\bibitem{} Mateo, M. 1996, in ``The formation of the Galactic halo: inside and out'', eds. H. Morrison and A. Sarajedini, ASP, 92, 343
\bibitem{} Mateo, M. et al. 1995, AJ, 109, 588
\bibitem{} Mateo, M. et al. 1996, ApJ, 458, L13
\bibitem{} McGee R.X. \& Newton L.M. 1986, Proc. Astron. Soc. Aust., 6, 471
\bibitem{} Moore, B. \& Davis, M. 1994, MNRAS, 270, 209
\bibitem{} Metcalf, B. \& Silk, J. 1996, ApJ, 464, 218
\bibitem{} Mathewson, D.H., Cleary, M.N., Murray, J.D.. 1974, ApJ, 190, 291
\bibitem{} Murai, T. \& Fujimoto, M. 1980, PASJ, 32, 581
\bibitem{} Navarro, J, Frenk, C., \& White, S. 1996, ApJ, 462, 563
\bibitem{} Oh K.S., Lin D.N.C., \& Aarseth S.J. 1995, ApJ, 442, 142
\bibitem{} Paczy\'nski, B. 1986, ApJ, 304, 1
\bibitem{} Payne-Gaposchkin, C.H. 1971, Smithsonian Contr. Ap., No. 13
\bibitem{} Quinn, P.J., Goodman, J. 1986, ApJ, 309, 472
\bibitem{} Quinn, P.J., Hernquist, L., Fullagar, D.P. 1993, ApJ, 403, 74
\bibitem{} Saha, A. 1985, ApJ, 289, 310
\bibitem{} Sahu, A. 1994, Nature, 370, 275
\bibitem{} Searle, L. \& Zinn, R. 1978, ApJ, 225, 357
\bibitem{} Smith, H. A. 1985, PASP, 97, 1053
\bibitem{} Toomre, A. \& Toomre, J. 1972, ApJ, 178, 623
\bibitem{} Tremaine, S. 1980, ``The structures and evolution of normal galaxies'', eds. S.M.Fall, D. Lynden-Bell, Cambridge University Press, Cambridge
\bibitem{} T\'oth, G., Ostriker, J.P. 1992, ApJ, 389, 5
\bibitem{} Unavane M., Wyse R.F.G. \& Gilmore G. 1996, MNRAS, 278, 727
\bibitem{} White, S.D.M. 1996, astro-ph/9602054
\bibitem{} Zhao, H.S. 1996, astro-ph/9606166

\end{thebibliography}
\end{document}